\documentclass[aps,preprint,floatfix,nofootinbib,showpacs]{revtex4-1}
\pdfoutput=1
\usepackage{graphicx,color}
\usepackage{hyperref}
\begin{document}

{\small
\begin{flushright}
CNU-HEP-14-03
\end{flushright} }

\title{Higgcision Updates 2014}

\renewcommand{\thefootnote}{\arabic{footnote}}

\author{
Kingman Cheung$^{1,2}$, Jae Sik Lee$^3$, and
Po-Yan Tseng$^1$}
\affiliation{
$^1$ Department of Physics, National Tsing Hua University,
Hsinchu 300, Taiwan \\
$^2$ Division of Quantum Phases and Devices, School of Physics, 
Konkuk University, Seoul 143-701, Republic of Korea \\
$^3$ Department of Physics, Chonnam National University, \\
300 Yongbong-dong, Buk-gu, Gwangju, 500-757, Republic of Korea
}
\date{\today}

\begin{abstract}
  During the 2014 Summer Conferences, both ATLAS and CMS
  Collaborations of the LHC experiments have demonstrated tremendous
  efforts in treatment of data and processing more data such that most
  data on signal strengths have improved; especially the diphoton and
  fermionic modes of both experiments.  Here in this note we perform an
  update to our previous model-independent Higgs precision analysis --
  Higgcision.  We found the followings: (i) the uncertainties on most
  couplings shrink about 10--20\%, (ii) the nonstandard
  (e.g. invisible) decay branching ratio of the Higgs boson is
  constrained to be less than $19\%$ if only the width 
  is allowed to
  vary, (iii) the gauge-Higgs coupling $C_v$ is constrained to be
  $0.94\,^{+0.11}_{-0.12}$, in which the uncertainty is reduced by
  about 10\%, and (iv) the standard model (SM) Higgs boson still provides
  the best fit to all the Higgs boson data, and compared to the previous
  results the SM Higgs boson now enjoys a higher $p$ value than the last year.
\end{abstract}

\maketitle

\section{Introduction}

It has been two years since a new particle was discovered at the Large
Hadron Collider (LHC) \cite{atlas,cms}. The initial data sets
indicated that it might be different from the standard model (SM)
Higgs boson. Nevertheless, after two more years of collecting data and
more painful scrutinizing the uncertainties and handling the
backgrounds, the data showed that it is more and more likely to be the
SM Higgs boson. Indeed, we showed in Ref.~\cite{higgcision} that the
SM Higgs boson provided the best fit to all the Higgs-boson data after
the Summer 2013. Similar results were obtained in a number of 
model-independent studies after Summer 2013 \cite{others}.

Most new Higgs boson results with improvements were presented in the 
ICHEP 2014 \cite{kado_ichep,david}. Especially the signal strength of the
diphoton decay channel of ATLAS has changed from
$1.6\pm 0.4$ to $1.17\pm 0.27$ \cite{atlas_aa_2014} and 
that of CMS from 
$0.78\,^{+0.28}_{-0.16}$ to $1.12\,^{+0.37}_{-0.32}$ \cite{cms_aa_2014}. 
Also, some updates were reported in the $ZZ$ \cite{atlas_zz_2014,cms_zz_2014},
the $WW$ \cite{mills_ichep,cms_ww_2014}, $b\bar b$ \cite{cms_bb_2014}, 
and $\tau^+ \tau^-$ \cite{cms_tau_2014} decay modes, as well as the 
$t\bar t H$ events \cite{tth_ichep,cms_bb_tth_2014} since 2013. 
There was also an overall update from the D\O \cite{tev_bb_2014}.
Here in this note
we present an update to our previous model-independent
Higgs precision analysis -- Higgcision \cite{higgcision}.

The organization of the paper 
is as follows. In the next section, 
we list all the improved Higgs boson data used in our global fits
and the updated fitting results are presented in Sec. III. 
We conclude in Sec. IV.

\section{Higgs Signal Strength Data}

The updated data on Higgs signal strengths are tabulated in 
Tables~\ref{aa} -- \ref{t5}.  We describe the notable differences between
the current data set and the one in Summer 2013.
\begin{itemize}
\item 
 $H\to \gamma\gamma$ has the most significant changes since the Summer 2013.
The ATLAS Collaboration updated their best-measured value 
$\mu_{ggH+ttH} = 1.6 \pm 0.4$ \cite{atlas_h_aa_2013}
to $\mu_{\rm inclusive}=1.17 \pm 0.27$ \cite{atlas_aa_2014}.  
The $\chi^2_{\rm SM}$ for the ATLAS $H\to \gamma\gamma$ channel reduces from 
$3.2$ to $0.96$. This data is now brought closer to the SM value.

\item The CMS $H\to\gamma\gamma$ data also entertains a dramatic change. 
It was the $\mu_{\rm untagged}=0.78\,^{+0.28}_{-0.26}$ \cite{cms_aa_2013}, 
and now it becomes $\mu_{ggH}= 1.12 \,^{+0.37}_{-0.32}$ \cite{cms_aa_2014}, 
which is now the most significant one
among the $\mu$'s for the diphoton channel. The underlying reason is 
believed to be better modeling of the diphoton background and calibration
of the photons.  This data is also brought closer to the SM value.

\item The ATLAS $H\to ZZ^*$ data increases from $1.5\pm 0.4$ 
\cite{atlas_com_2013} to $1.66 \,^{+0.45}_{-0.38}$ \cite{atlas_zz_2014}, 
such that the $\chi^2_{\rm SM}$ jumps from
$1.6$ to $3.0$. The CMS $H\to ZZ^*$ stays about the same 
\cite{cms_zz_2013,cms_zz_2014}. The $ZZ^*$ 
channel becomes more important in the overall fitting.

\item The CMS $H\to WW^*$ data shows some improvements 
\cite{cms_ww_2013,cms_ww_2014}. While the
$\mu$ (0/1 jet) stays about the same, the $\mu_{VBF}$ and $\mu_{VH}$ show
positive central values now. Also, a new $\mu_{WH} (3\ell 3\nu)$ is now
available.  The ATLAS $H\to WW^*$ remains the same 
\cite{atlas_com_2013,mills_ichep}.

\item Both ATLAS and CMS show improvements in $H\to b\bar b$ channel
because a close-to-full set of data was analyzed. Both show the 
VH tag and ttH tag results
\cite{cms_bb_2014,atlas_bb_2013,tth_ichep}. See Table~\ref{bb}.

\item
The ATLAS $H\to \tau\tau$ channel has analyzed the full-set of data and 
shows some improvements: both central values
are close to 1 and the size of errors is reduced \cite{atlas_h_tau_2013}.
The CMS now separately
reported $0$-jet and $1$-jet categories. The size of errors slightly improves
\cite{cms_tau_2014}.

\item 
The deviations from the SM for each Higgs decay channel in terms of 
$\chi^2_{\rm SM}$ are listed in the last column in Tables~\ref{aa} -- \ref{t5}.
Previously, it is largely dominated by the $\gamma\gamma$ channel but now
its significance is somewhat reduced and the $ZZ^*$ becomes more important.

\item The total $\chi^2_{\rm SM}$/d.o.f. for the SM is now $16.76/29$, 
compared to the previous one $18.92/22$. Substantial improvement for the
SM can be seen.  The $p$ value for the SM increases from $0.65$ to $0.966$.

\item
We also note that the Higgs boson mass measurement at ATLAS now gives
$M_H=125.4 \pm 0.4$ GeV
($H \to \gamma\gamma$) \cite{atlas_aa_2014}
and 
$124.51 \pm 0.52~({\rm stat}) \pm 0.06~({\rm syst})$ GeV ($H \to ZZ^*$)
\cite{kado_ichep} 
with the smaller difference and noticeable improvement on systematics.
The corresponding CMS values are
$M_H=124.70 \pm 0.31~({\rm stat}) \pm 0.15~({\rm syst})$ GeV 
($H \to \gamma\gamma$) and 
$125.6 \pm 0.4~({\rm stat}) \pm 0.2~({\rm syst})$ GeV ($H \to ZZ^*$)
\cite{david}.
It is interesting to note that the ATLAS $H\to \gamma\gamma$ ($H \to ZZ^*$) 
value  is similar  in size to the CMS $H \to ZZ^*$ ($H\to \gamma\gamma$)
one.
\end{itemize}

\section{Fits}

The formalism, notation, and convention follow closely 
our earlier paper \cite{higgcision}.  We restate the notation of
the coupling parameters that are used here.  The normalized 
Yukawa couplings are
\begin{eqnarray}
&&
C_u^S=g^S_{H\bar uu}\,, \ \
C_d^S=g^S_{H\bar dd}\,, \ \
C_\ell^S=g^S_{H\bar ll}\,; \ \
C_v=g_{_{HVV}}\,; \nonumber \\
&&
C_u^P=g^P_{H\bar uu}\,, \ \
C_d^P=g^P_{H\bar dd}\,, \ \
C_\ell^P=g^P_{H\bar ll}\,,
\end{eqnarray}
where the superscripts ``S'' and ``P'' denote scalar and pseudoscalar
couplings. In the SM, all scalar Yukawa couplings and $C_v$ equal 1 while
the pseudoscalar ones equal 0.
Here we also assume generation independence and custodial 
symmetry between the $W$ and $Z$ bosons. 
In the fits, the deviations due to additional particles running in the 
triangular loops of the $H\gamma\gamma$ and $Hgg$ vertices are given by
\begin{equation}
\Delta S^\gamma\,,  \ \ \ \Delta P^\gamma\,; \ \ \Delta S^g\,, \ \ 
\Delta P^g\,.
\end{equation}
In the SM, these factors are 0.
Finally,
the additional contribution to the width of the Higgs boson,
$\Delta \Gamma_{\rm tot}$, is also used.

The labeling of the fits is as follows: CPC denotes CP conserving, CPV denotes
CP violating, and the number after CPC or CPV denotes the number of
varying parameters.

\subsection{CP Conserving fits}

In the CP conserving fits, we have set all pseudoscalar couplings and
form factors to be zero:
\begin{equation}
C^P_{u,d,\ell}=\Delta P^{g,\gamma}=0\,,
\end{equation}
while varying
\begin{equation}
C^S_{u,d,\ell}, \ \ \
C_v, \ \ \
\Delta S^{g,\gamma}, \ \ \
\Delta\Gamma_{\rm tot}.
\end{equation}

\subsubsection{CPC1: vary only $\Delta \Gamma_{\rm tot}$ while keeping
$C^S_u=C^S_d=C^S_\ell=C_v = 1$ and 
$\Delta S^\gamma = \Delta S^g = 0$}
\begin{figure}[h!]
\centering
\includegraphics[width=4in,angle=270]{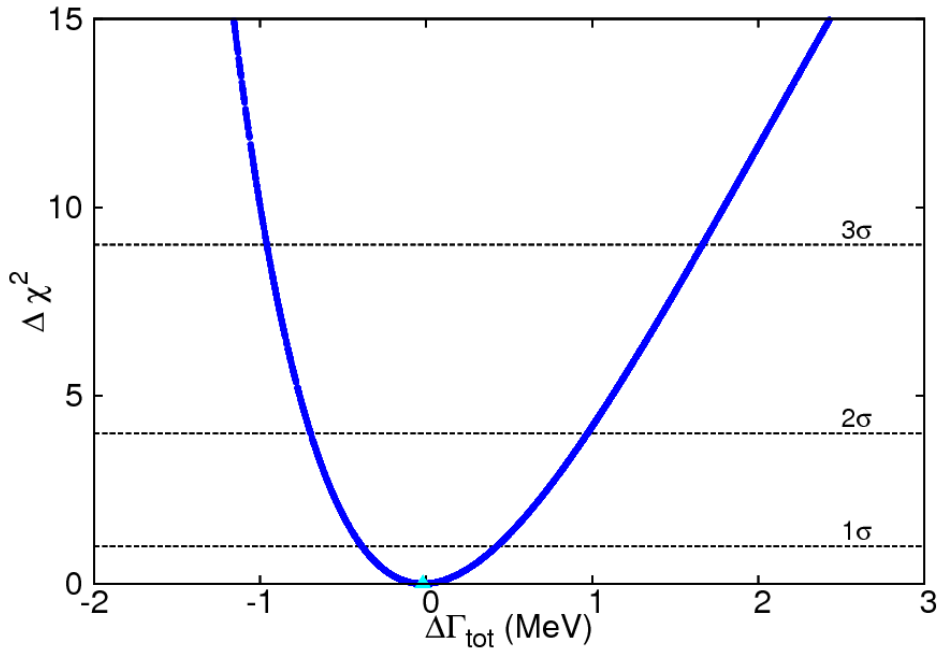}
\caption{\small \label{width-chi}
Variation of $\Delta \chi^2$ versus $\Delta \Gamma_{\rm tot}$ 
in the CPC1 case.}
\end{figure}

The $\Delta \Gamma_{\rm tot}$ can account for additional decay modes of
the observed Higgs boson, in particular the celebrated invisible decay
mode into hidden-sector particles, dark matter, etc.  The $1$-$\sigma$
fitted value is shown in the second column of 
Table~\ref{bestfit}.  
The variation of chi-square around the minimum chi-square point is 
shown in Fig.~\ref{width-chi}.
We found that
$\Delta \Gamma_{\rm tot}$ alone does not improve the chi-square and the
central value is consistent with 0. The 95\% allowed range for
$\Delta \Gamma_{\rm tot}$ is 
\begin{equation}
\Delta \Gamma_{\rm tot} = -0.020\;^{+0.97}_{-0.66} \; {\rm MeV}\;.
\end{equation}
Comparing to the previous value $0.10\,^{+1.11}_{-0.74}$ MeV, the upper
error improves by more than 10\%. Thus, the 95\% CL upper limit for
$\Delta \Gamma_{\rm tot}$ improves to $0.97$ MeV.  Therefore,
the nonstandard branching ratio of the Higgs boson is constrained to be
\begin{equation}
 B(H \to {\rm nonstandand}) < 19\% \;,
\end{equation}
which shows a 14\% improvement from the previous value of 22\%.
Note that such a stringent bound can be relaxed when other parameters
are allowed to vary in the fits. For example, in the CPC3 fit below the
1-$\sigma$ range for $\Delta \Gamma_{\rm tot}$ is $0.39 \,^{+1.13}_{-0.76}$,
which can easily translate into a branching ratio of order 50\%.
This is consistent with a direct search for invisible decay of the
Higgs boson \cite{invisible}.  

\subsubsection{CPC2: vary $\Delta S^\gamma$ and $\Delta S^g$ while keeping
$C_u^S=C_d^S=C_\ell^S=C_v = 1$}
This case can account for additional electrically charged particles or
colored particles running in the triangular loops of 
$H\gamma\gamma$ and $Hgg$, respectively, while the Yukawa and gauge-boson
couplings take on SM values. The best fit value for CPC2 is shown in the
third column of Table~\ref{bestfit}. The central value of $\Delta S^\gamma
= -0.72$ is to increase the form factor $S^\gamma_{\rm SM} = -6.64$ 
\cite{higgcision} by about 11\% while the central value of
$\Delta S^g = -0.009$ decreases form factor $S^g_{\rm SM} = 0.64$ by about 
1\%.
Overall, the diphoton rate, which is proportional to
$|S^\gamma|^2 |S^g|^2$ increases by about 20\% so as to match the
central values of the ATLAS and CMS data.
Although the total chi-square reduced by $1.0$ units, the $p$ value
of this CPC2 fit is almost the same as the SM $p$ value. 

\subsubsection{CPC3: vary $\Delta S^\gamma$, $\Delta S^g$, and
$\Delta \Gamma_{\rm tot}$  while keeping
$C_u^S=C_d^S=C_\ell^S=C_v = 1$}
We have one more varying parameter, $\Delta \Gamma_{\rm tot}$,
than the previous case. The central values and uncertainties 
for $\Delta S^\gamma$ and $\Delta S^g$ are similar to the previous case. 
No improvement from the previous one is seen, as shown in the fourth 
column of Table~\ref{bestfit}. 
The 1-$\sigma$ range for $\Delta \Gamma_{\rm tot}$ is $0.39 \, ^{+1.13}_{-0.76}$
MeV. In this 3-parameter fit, the constraint on $\Delta \Gamma^{\rm tot}$ is
much relaxed than the CPC1 fit, because of marginalizing the other two
parameters. Such a large uncertainty in $\Delta \Gamma_{\rm tot}$ would
give a nonstandard decay branching ratio of order $O(50)$\%
for the Higgs boson.

\subsubsection{CPC4: vary 
$C^S_u$, $C^S_d$, $C^S_\ell$, $C_v$  while keeping
$\Delta S^\gamma = \Delta S^g = 0$}
In this fit, only the Yukawa and gauge couplings are allowed to vary,
which will in turn modify the form factors $S^\gamma$ and $S^g$,
even though $\Delta S^\gamma$ and $\Delta S^g$ are fixed at zero. 
The result is similar to the one in Summer 2013, as shown in the fifth 
column of Table~\ref{bestfit}. The gauge coupling is constrained to
be $0.98\,^{+0.10}_{-0.11}$ compared to the previous one at 
$1.04\,^{+0.12}_{-0.14}$. We can see some improvement in the uncertainty
of order 20\%. The uncertainties in Yukawa couplings are about the same.

\begin{figure}[h!]
\includegraphics[width=4.5in,angle=270]{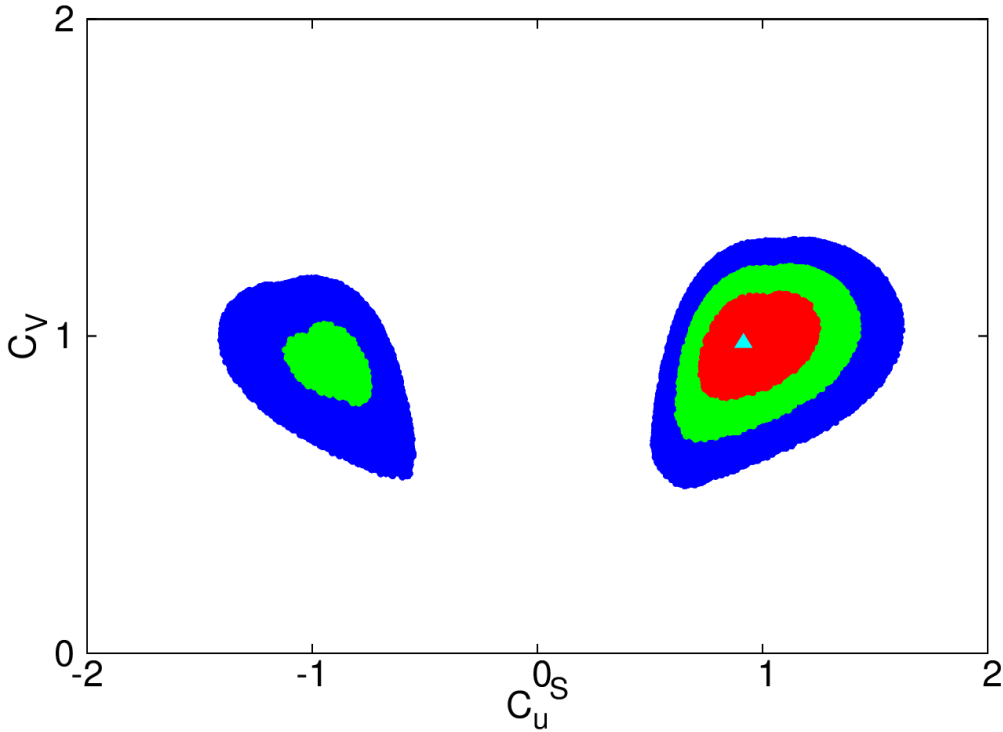}
\caption{\small \label{cpc4-fig}
The confidence-level regions in the plane of $(C_u^S,\;C_v)$ 
of the CPC4 fit by varying $C_u^S$, $C_d^S$,  $C_l^S$, and $C_v$ 
while keeping $\Delta S^\gamma = \Delta S^g = \Delta \Gamma_{\rm tot} = 0$. 
The contour regions shown are for $\Delta \chi^2 \leq 2.3$ (red), 5.99 (green), 
and 11.83 (blue) above the minimum, which correspond to confidence levels of 
68.3\%, 95\%, and 99.7\%, respectively. The best-fit point is denoted by 
the triangle.
}\end{figure}

The sign of the top-Yukawa coupling $C_u^S$ is the most nontrivial one
in this case because the coefficients of the $W$ and top contributions to
the form factor $S^\gamma$ come in comparable size but of opposite sign.
The top-quark contribution tends to offset parts of the $W$ contribution
when $C_u^S > 0$ but enhance when $C_u^S < 0$. Note that the contributions
from the bottom quark and tau lepton are relatively much smaller.
The result shown in Fig.~3a of Ref.~\cite{higgcision}, as of Summer 2013,
indicated that positive $C_u^S$ around $0.8$ was preferred but
around $-0.9$ was still allowed at 95\% CL. Now we show the corresponding
plot based on the most updated data in Summer 2014 in Fig.~\ref{cpc4-fig}.
We can see that the value of $C_u^S$ around $0.9$ is more preferred than
before while the island around $-0.9$ diminished.

\subsubsection{CPC6: vary 
$C_u^S$, $C_d^S$, $C_\ell^S$, $C_v$, $\Delta S^\gamma$, $ \Delta S^g $}
This CPC case has the most varying parameters.  We found that there
are a few sets of degenerate solutions. We first discuss the one with
a smaller $|\Delta S^g|$ because the new colored particles are likely to
be heavy, and a positive top-Yukawa coupling as a more conventional choice
though it is not necessarily the case. The result is shown in the last
column of Table~\ref{bestfit}.
Compared to the result in Summer 2013 \cite{higgcision},
where we have $C_u^S=0.00\pm 1.13$, the most notable
difference is that the top-Yukawa $C_u^S$ takes on a nonzero value,
and correspondingly the $\Delta S^\gamma$ turns negative or positive
due to its correlation to $C_u^S$. 
Specifically, we observe that the $C_u^S=0$ hypothesis 
has been ruled out at 68.3\% CL.

\begin{figure}[h!]
\includegraphics[width=2.4in,angle=270]{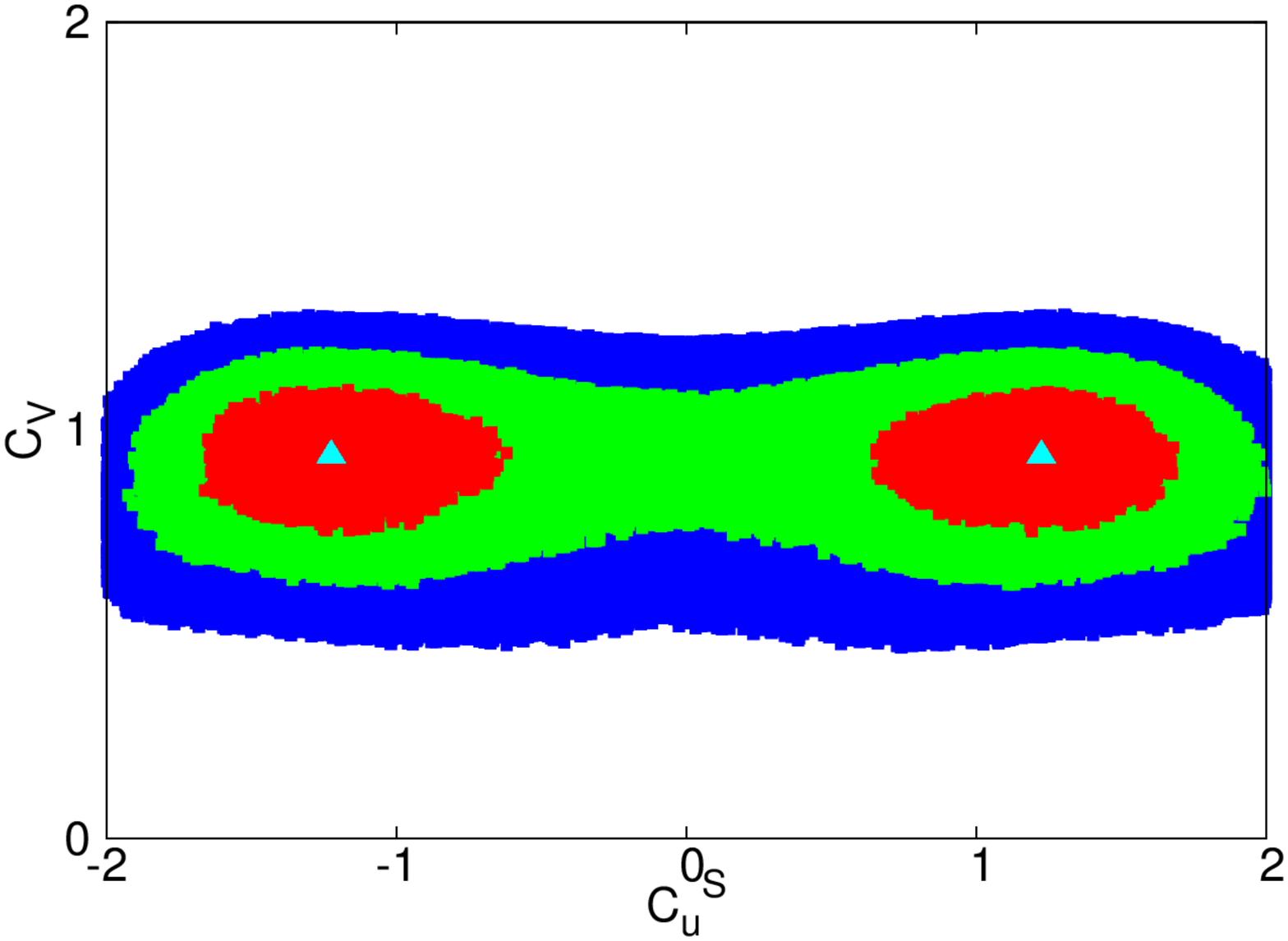}
\includegraphics[width=2.4in,angle=270]{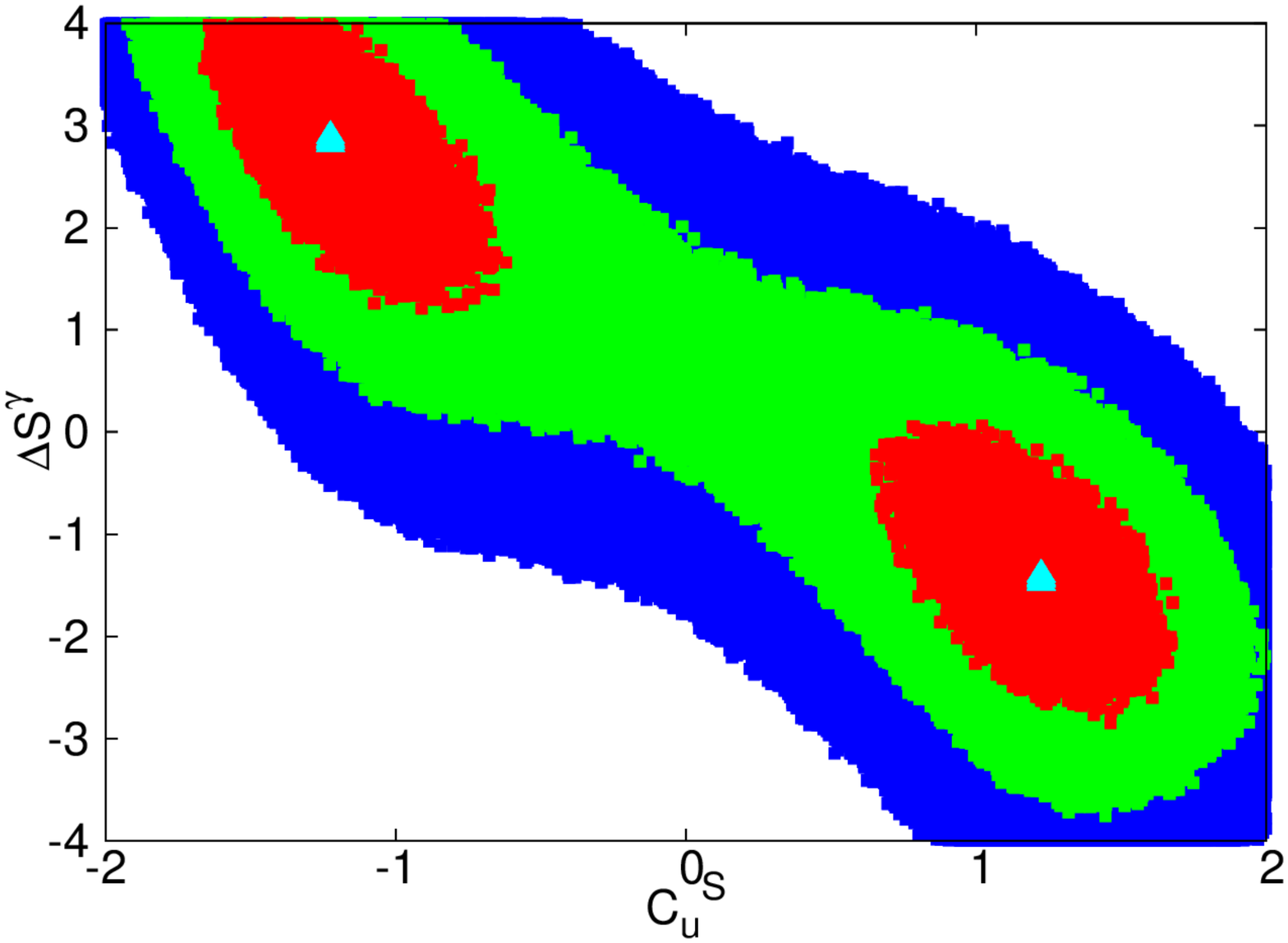}
\includegraphics[width=2.4in,angle=270]{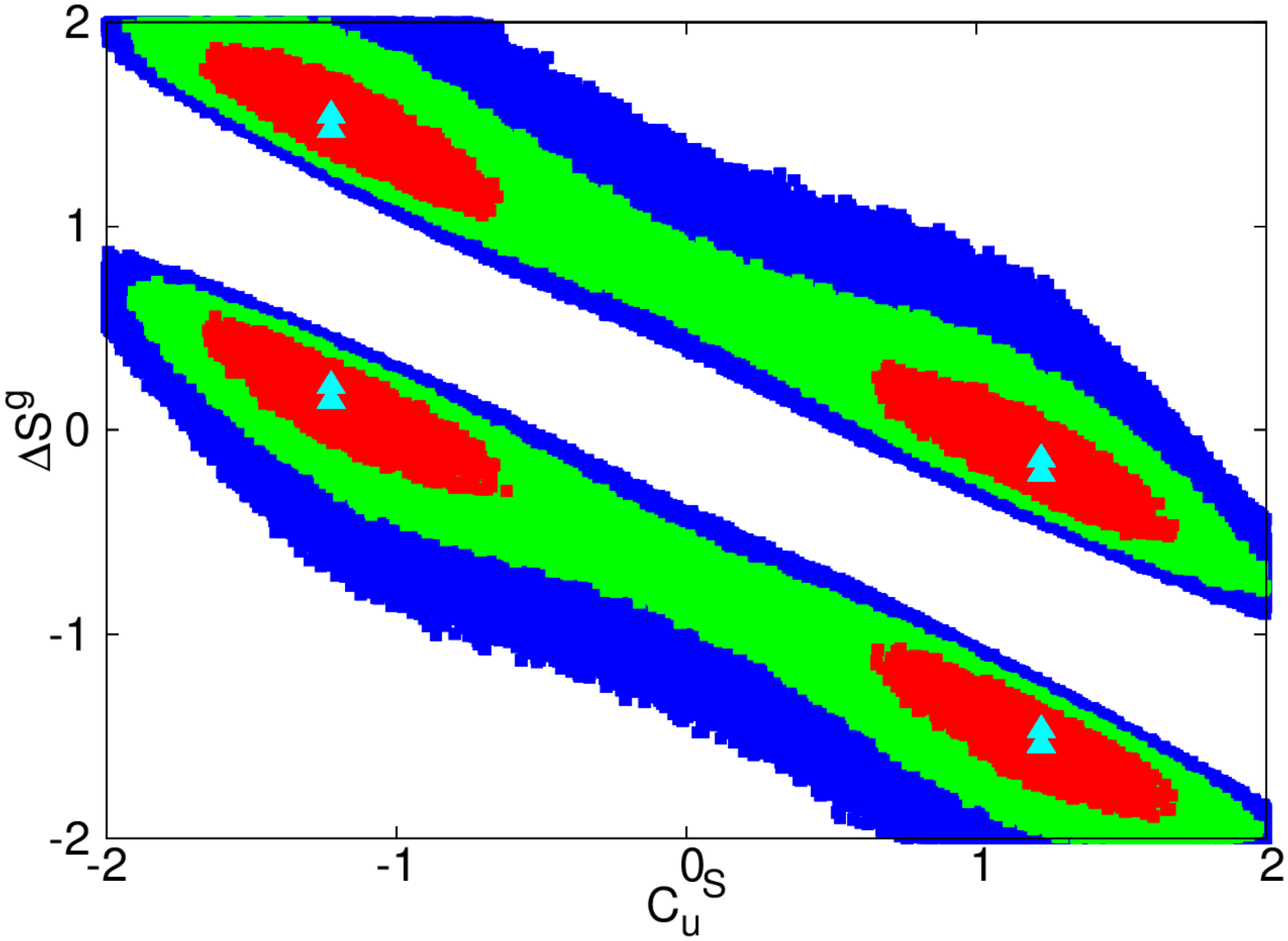}
\caption{\small \label{cpc6-fig}
The confidence-level regions in the plane of $(C_u^S,\;C_v)$, 
$(C_u^S,\; \Delta S^\gamma)$, an $(C_u^S,\; \Delta S^g)$
 of the CPC6 fit by varying $C_u^S$, $C_d^S$,  $C_l^S$, $C_v$,
$\Delta S^\gamma$, and $\Delta S^g$.
The contour regions shown are for $\Delta \chi^2 \leq 2.3$ (red), 5.99 (green), 
and 11.83 (blue) above the minimum, which correspond to confidence levels of 
68.3\%, 95\%, and 99.7\%, respectively. The best-fit points are denoted by 
the triangles.
}\end{figure}

As we have just said, there are a few sets of degenerate solutions. They
all give the same chi-square and thus the same $p$ value,
as  shown in Table~\ref{bestfit6} and Fig.~\ref{cpc6-fig}.
The degenerate solutions arise because the diphoton
rate is proportional to $|S^\gamma|^2 |S^g|^2$. Numerically, $S^\gamma$ and 
$S^g$ are given by \cite{higgcision}
\begin{eqnarray}
S^\gamma &\simeq & -8.35 \, C_v + 1.76\,  C_u^S + \Delta S^\gamma \nonumber \\
S^g &\simeq & 0.69\,  C_u^S + \Delta S^g \nonumber
\end{eqnarray}
In Table~\ref{bestfit6}, the value of $C_v =0.94$.  
With $|\Delta S^\gamma|\leq 4$,
four categories of solutions
exist for $(C_u^S,\, \Delta S^\gamma,\, \Delta S^g) \approx 
(1.22,\, -1.4,\,-0.2), \; (1.22,\, -1.4,\,-1.5), \; (-1.23,\, 2.8,\,0.2) ,\;
(-1.23,\, 2.8,\,1.5)$. In each category, we find four solutions corresponding
to the four combinations of the signs of $C_{d,\ell}^S$.

\subsubsection{CPC N2: vary only 
$C^S_u$ and $C_v$  while keeping others fixed}
In all previous CPC fits, we find that the most effective parameters
are $C_v$, $C_u^S$, and $\Delta S^\gamma$ but less on $\Delta S^g$.  We first
look at $C_u^S$ and $C_v$ in this special two-parameter fit CPC~N2.
It is listed in the second column of the lower panel in Table~\ref{bestfit}.
Both $C_u^S$ and $C_v$ are close to the SM values with less than 
10\% uncertainty. 

\subsubsection{CPC N3: vary only 
$C^S_u$, $C_v$, and $\Delta S^\gamma$  while keeping others fixed}
The result is shown in third column of the lower panel in Table~\ref{bestfit},
and it is similar to the CPC2 case with $C_v$ and $C_u^S$ close to 1
and $\Delta S^\gamma$ close to $-1$.

\subsubsection{CPC N4: vary only 
$C^S_u$, $C_v$, $\Delta S^\gamma$, and $\Delta S^g$ while keeping others fixed}
The result is shown in the last four columns of the lower panel 
in Table~\ref{bestfit}, and very similar to that of the CPC6 case. 
There are four sets of solutions.
They have a higher
$p$-value than in the CPC6 case since we are taking $C_d^S=C_\ell^S=1$, 
which does not much affect our fits to the current data.

\subsection{CP violating fits}
We devote this section to including the pseudoscalar top-Yukawa coupling
$C_u^P$, 
and the pseudoscalar contributions $\Delta P^\gamma$ and $\Delta P^g$.

\subsubsection{CPV3: vary $C_u^S$, $C_u^P$ and $C_v$}
\begin{figure}[h!]
\centering
\includegraphics[width=4.5in,angle=270]{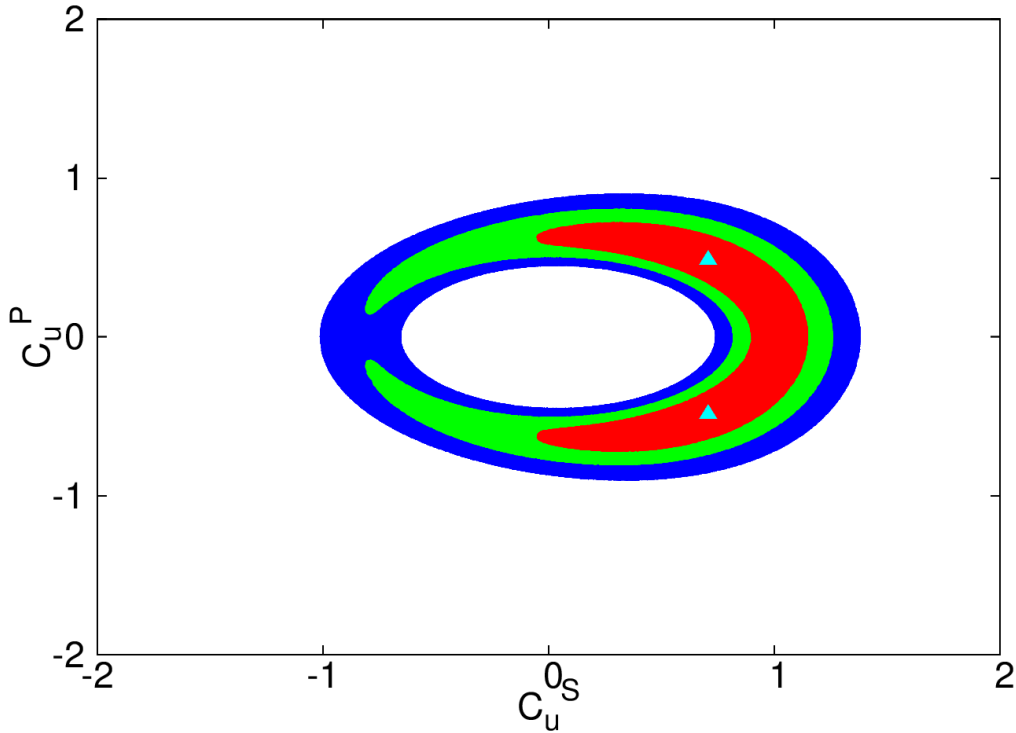}
\caption{\small \label{cpv3-fig}
The confidence-level regions of the fit by varying the scalar Yukawa 
couplings $C_u^S$ and $C_v$ , and the pseudoscalar Yukawa couplings $C_u^P$; 
while keeping others at the SM values. 
The description of contour regions is the same as in Fig.~\ref{cpc4-fig}.
}
\end{figure}

We have shown in Ref.~\cite{higgcision} that $C_u^S$ and $C_u^P$ 
satisfy an elliptical equation, so that the allowed regions
display elliptical shapes, as shown in Fig.~\ref{cpv3-fig}.
We also find that the size of uncertainties decreases slightly, as
shown in the second and third columns of Table~\ref{cbestfit}.

\subsubsection{CPV4: vary $\Delta S^\gamma$, $\Delta S^g$,  
$\Delta P^\gamma$, and $\Delta P^g$ }
The result is shown in the last two columns in Table~\ref{cbestfit},
and more or less similar to the result of Summer 2013 \cite{higgcision}
taking account of the large errors of $\Delta S^\gamma$ and $\Delta P^\gamma$. 

\subsubsection{CPV N4: vary $C_u^S$, $C_u^P$, $C_v$, $\Delta S^\gamma$;
CPV N5: vary $C_u^S$, $C_u^P$, $C_v$, $\Delta S^\gamma$, $\Delta P^\gamma$}
The results are shown in the second and third columns of 
Table~\ref{cbestfitn}. We observe that the pseudoscalar couplings
$C_u^P$ and $\Delta P^\gamma$ are consistent with zero, and thus 
the fits are otherwise  the same as CPC N3.

\subsubsection{CPV N6: vary $C_u^S$, $C_u^P$, $C_v$, $\Delta S^\gamma$,
$\Delta P^\gamma$, $\Delta S^g$;
CPV N7: vary $C_u^S$, $C_u^P$, $C_v$, $\Delta S^\gamma$, $\Delta P^\gamma$,
$\Delta S^g$, $\Delta P^g$}
The results are shown in the last two  columns of 
Table~\ref{cbestfitn}. We observe that the pseudoscalar couplings
$C_u^P$, $\Delta P^\gamma$, and $\Delta P^g$ are close to zero within 
uncertainties, and thus 
the fits are otherwise the same as CPC N4. 

\section{Discussion}

\begin{figure}[h!]
\centering
\includegraphics[width=4.5in]{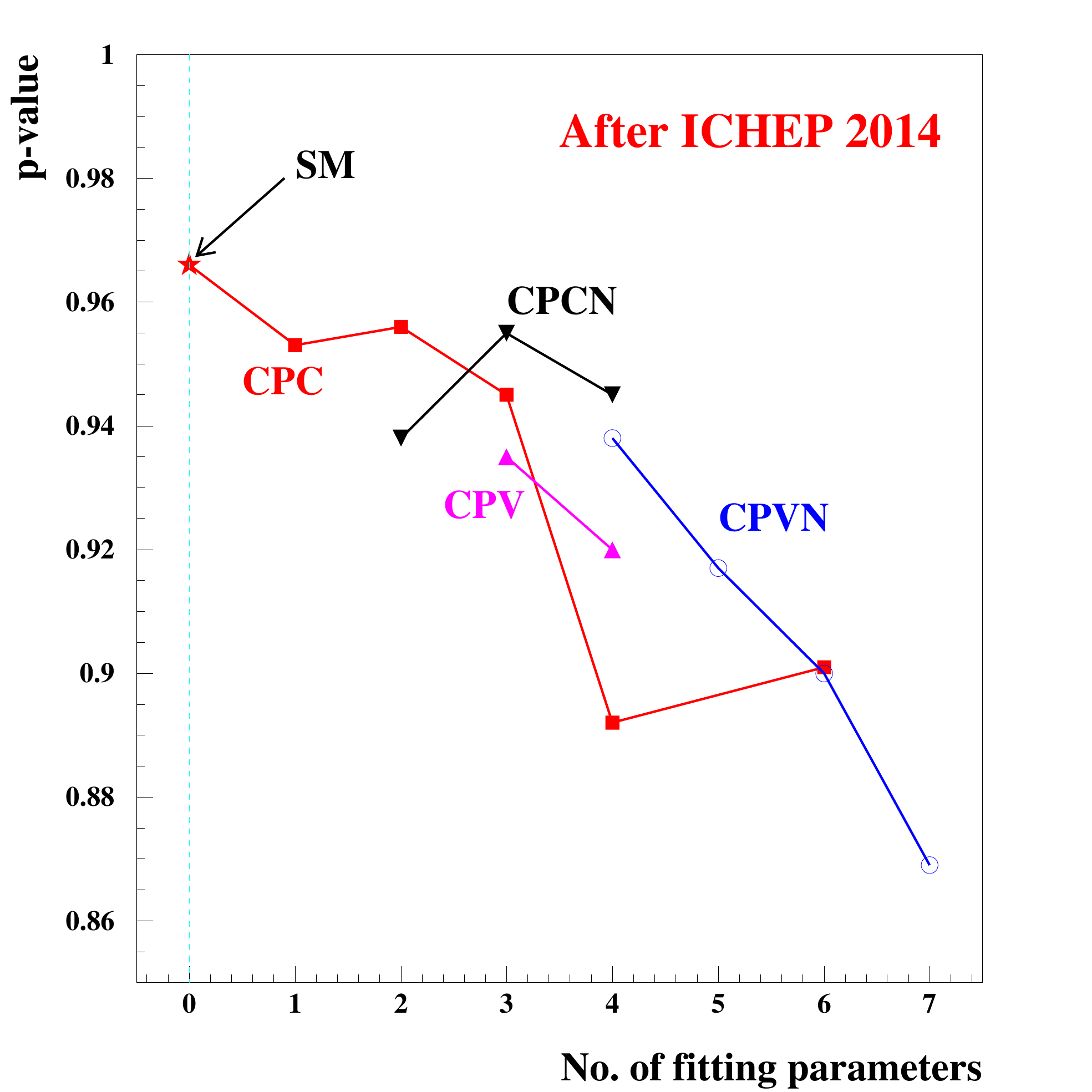}
\caption{\small \label{pval}
The $p$-values for various fits considered in this work, including
CP-conserving (CPC) and CP-violating (CPV) ones.
}
\end{figure}

The most significant updates between Summer 2013 and 2014 are the diphoton
signal strengths reported by both ATLAS and CMS, and also the full set
of data was analyzed for fermionic channels. Although more decay channels
are analyzed, e.g., $WH \to 3\ell 3\nu$ and $ttH\to tt \gamma\gamma$, 
the uncertainties are still too large to give any impact to the fits.
The $p$ values for all the fits are plotted in Fig.~\ref{pval}. The SM
still provides the best fit to the whole set of data and enjoys a
substantial better $p$ value than the last year.

We offer the following comments before we close.
\begin{enumerate}
\item
 The most constrained coupling is the Higgs-gauge coupling. In various
fits, it is constrained to the range $0.93 - 1.00$ with about $7 - 12\%$
uncertainties. The uncertainties are reduced by about 10\%.

\item
The CPC top-Yukawa coupling $C_u^S$ is now more preferred to be
positive in those fits
with $\Delta S^\gamma$ and $\Delta S^g$ fixed at zero. This is because
the top contribution with $C_u^S>0$ can cancel a part of the $W$ contribution
such that the prediction is close to the SM value and thus the data.

\item 
The data cannot rule out the pseudoscalar couplings, as shown in 
Fig.~\ref{cpv3-fig}, and only a combination of $C_u^S$ and $C_u^P$ is
constrained in the form of an elliptical equation.

\item 
One of the most useful implications is the invisible decay branching ratio
of the Higgs boson.  We obtain $B(H \to {\rm nonstandard}) < 19\%$.

\item
The possibility of having vanishing $C_u^S$ has been ruled out at 68.3\% CL.

\end{enumerate}

\section*{Acknowledgment}  
This work was supported the National Science
Council of Taiwan under Grants No. NSC 102-2112-M-007-015-MY3.
J.S.L. was supported by
the National Research Foundation of Korea (NRF) grant
(No. 2013R1A2A2A01015406).
This study was also
financially supported by Chonnam National University, 2012.
J.S.L thanks National Center for Theoretical Sciences (Hsinchu, Taiwan) for the 
great hospitality
extended to him while this work was being performed.

\begin{table}[thb!]
\caption{\small \label{aa}
Data on signal strengths of $H\rightarrow \gamma \gamma$
by the ATLAS and CMS, and at the Tevatron after ICHEP 2014.
The luminosity updates at 8 TeV are shown in the parenthesis.
The percentages of each production mode in each data are given.
The $\chi^2$ of each data with respect to the SM is shown in the last column.
The sub-total $\chi^2$ of this decay mode is shown at the end.
}
\begin{ruledtabular}
\begin{tabular}{cccccccr}
Channel & Signal strength $\mu$ & $M_H$(GeV) & \multicolumn{4}{c}{Production mode}  & $\chi^2_{\rm SM}$(each)\\
        & c.v $\pm$ error       &            & ggF & VBF & VH & ttH \\
\hline
\multicolumn{8}{c}
{ATLAS (4.5$fb^{-1}$ at 7TeV + 20.3$fb^{-1}$ at 8TeV): page 29 of \cite{atlas_aa_2014} (Aug. 2014)}\\
\hline
$\mu_{ggH}$ & $1.32 \pm 0.38$ & 125.40 & 100\% & - & - & - & 0.71 \\
$\mu_{VBF}$ & $0.8 \pm 0.7$ & 125.40 & - & 100\% & - & - & 0.08 \\
$\mu_{WH}$ & $1.0 \pm 1.6$ & 125.40 & - & - & 100\% & - & 0.00 \\
$\mu_{ZH}$ & $0.1^{+3.7}_{-0.1}$ & 125.40 & - & - & 100\% & - & 0.06 \\
$\mu_{ttH}$ & $1.6^{+2.7}_{-1.8}$ & 125.40 & - & - & - & 100\% & 0.11 \\
\hline
\multicolumn{7}{c}
{CMS (5.1$fb^{-1}$ at 7TeV + 19.7$fb^{-1}$ at 8TeV): Fig. 24 of \cite{cms_aa_2014} (July 2014) }\\
\hline
$\mu_{ggH}$ & $1.12^{+0.37}_{-0.32}$ & 124.70 & 100\% & - & - & - & 0.14 \\
$\mu_{VBF}$ & $1.58^{+0.77}_{-0.68}$ & 124.70 & - & 100\% & - & - & 0.73 \\
$\mu_{VH}$ & $-0.16^{+1.16}_{-0.79}$ & 124.70 & - & - & 100\% & - & 1.00 \\
$\mu_{ttH}$ & $2.69^{+2.51}_{-1.81}$ & 124.70 & - & - & - & 100\% & 0.87 \\
\hline
\multicolumn{7}{c}
{Tevatron (10.0$fb^{-1}$ at 1.96TeV): page 32 of \cite{tev} (Nov. 2012)}\\
\hline
Combined & $6.14^{+3.25}_{-3.19}$ & 125 & 78\% & 5\% & 17\% & - & 2.60 \\
\hline
&&&&&&& subtot: 6.30 

\end{tabular}
\end{ruledtabular}
\end{table}

\begin{table}[thb!]
\caption{\small \label{zz}
The same as Table~\ref{aa} but for $H\rightarrow Z Z^{(\ast)}$}
\begin{ruledtabular}
\begin{tabular}{cccccccr}
Channel & Signal strength $\mu$ & $M_H$(GeV) & \multicolumn{4}{c}{Production mode}  & $\chi^2_{\rm SM}$(each)\\
        & c.v $\pm$ error       &            & ggF & VBF & VH & ttH \\
\hline
\multicolumn{8}{c}
{ATLAS (4.5$fb^{-1}$ at 7TeV + 20.3$fb^{-1}$ at 8TeV): page 18 of \cite{atlas_zz_2014} (June 2014), page 18 of \cite{kado_ichep}}\\
\hline
Inclusive & $1.66^{+0.45}_{-0.38}$ & 124.51 & 87.5\% & 7.1\% & 4.9\% & 0.5\% & 
3.02 \\
\hline
\multicolumn{7}{c}
{CMS (5.1$fb^{-1}$ at 7TeV + 19.7$fb^{-1}$ at 8TeV): abstract of \cite{cms_zz_2014} (Dec. 2013) }\\
\hline
Inclusive & $0.93^{+0.29}_{-0.25}$ & 125.6 & 87.5\% & 7.1\% & 4.9\% & 0.5\% & 
0.06 \\
\hline
&&&&&&& subtot: 3.07 

\end{tabular}
\end{ruledtabular}
\end{table}

\begin{table}[thb!]
\caption{\small \label{ww}
The same as Table~\ref{aa} but for $H\rightarrow W W^{(\ast)}$}
\begin{ruledtabular}
\begin{tabular}{cccccccr}
Channel & Signal strength $\mu$ & $M_H$(GeV) & \multicolumn{4}{c}{Production mode} & $\chi^2_{\rm SM}$(each) \\
        & c.v $\pm$ error       &            & ggF & VBF & VH & ttH \\
\hline
\multicolumn{7}{c}
{ATLAS (4.6$fb^{-1}$ at 7TeV + 20.7$fb^{-1}$ at 8TeV): page 10 of \cite{mills_ichep} (July 2014)}\\
\hline
Inclusive & $0.99\pm 0.30$ & 125 & 87.5\% & 7.1\% & 4.9\% & 0.5\% & 0.00 \\
\hline
\multicolumn{7}{c}
{CMS (4.9 $fb^{-1}$ at 7TeV + 19.4$fb^{-1}$ at 8TeV):Fig. 23 of \cite{cms_ww_2014} (Dec. 2013) }\\
\hline
0/1 jet & $0.74^{+0.22}_{-0.20}$ & 125.6 & 97\% & 3\% & - & - & 1.40 \\
VBF tag & $0.60^{+0.57}_{-0.46}$ & 125.6 & 17\% & 83\% & - & - & 0.49 \\
VH tag ($2l2\nu 2j$) & $0.39^{+1.97}_{-1.87}$ & 125.6 & - & - & 100\% & - & 0.10
 \\
WH tag ($3l3\nu$) & $0.56^{+1.27}_{-0.95}$ & 125.6 & - & - & 100\% & - & 0.12 \\
\hline
\multicolumn{7}{c}
{Tevatron (10.0$fb^{-1}$ at 1.96TeV): Page32 of \cite{tev} (Nov. 2012)}\\
\hline
Combined & $0.85^{+0.88}_{-0.81}$ & 125 & 78\% & 5\% & 17\% &  - & 0.03 \\
\hline
&&&&&&& subtot: 2.14 
\end{tabular}
\end{ruledtabular}
\end{table}
\begin{table}[thb!]
\caption{\small \label{bb}
The same as Table~\ref{aa} but for $H\rightarrow b\bar{b}$}
\begin{ruledtabular}
\begin{tabular}{cccccccr}
Channel & Signal strength $\mu$ & $M_H$(GeV) & \multicolumn{4}{c}{Production mode}  & $\chi^2_{\rm SM}$(each)\\
        & c.v $\pm$ error       &            & ggF & VBF & VH & ttH \\
\hline
\multicolumn{8}{c}
{ATLAS (4.7(4.5)${\rm fb}^{-1}$ at 7TeV + 20.3${\rm fb}^{-1}$ at 8TeV)
  Fig.19 of \cite{atlas_bb_2013} (July 2013), page 18 of \cite{tth_ichep}}\\

\hline
VH tag & $0.2^{+0.7}_{-0.6}$ & 125.5 & - & - & 100\% & - & 1.31 \\
ttH tag & $1.8^{+1.66}_{-1.57}$ & 125.4 & - & - & - & 100\% & 0.26 \\
\hline
\multicolumn{8}{c}
{CMS (5.1${\rm fb}^{-1}$ at 7TeV + 18.9${\rm fb}^{-1}$ at 8TeV)
 \cite{cms_bb_2014} (Oct. 2013), (19.5${\rm fb}^{-1}$ at 8TeV)\cite{cms_bb_tth_2014} (July 2014) }\\
\hline
VH tag & $1.0 \pm 0.5$ & 125 & - & - & 100\% & - & 0.00 \\
ttH tag & $0.67^{+1.35}_{-1.33}$ & 125 & - & - & - & 100\% & 0.06\\
\hline
\multicolumn{7}{c}
{Tevatron (10.0${\rm fb}^{-1}$ at 1.96TeV): \cite{tev_bb_2014} }\\
\hline
VH tag & $1.59^{+0.69}_{-0.72}$ & 125 & - & - & 100\% & - & 0.67 \\
\hline
&&&&&&& subtot: $2.30$
\end{tabular}
\end{ruledtabular}
\end{table}
\begin{table}[thb!]
\caption{\small \label{t5}
The same as Table~\ref{aa} but for $H\rightarrow \tau \tau$. 
The correlation for the $\tau\tau$ data of ATLAS is $\rho=-0.51$. }
\begin{ruledtabular}
\begin{tabular}{c c c cccc r}
Channel & \multicolumn{1}{c}{Signal strength $\mu$} & $M_H$(GeV) & 
\multicolumn{4}{c}{Production mode} & \multicolumn{1}{c}{$\chi^2_{\rm SM}$(each)}
 \\
  &   && ggF & VBF & VH & ttH &  \\
\hline
\multicolumn{8}{c}
{ATLAS (20.3${\rm fb}^{-1}$ at 8TeV): page 28 of \cite{atlas_h_tau_2013} (Nov. 2013)}\\
\hline
$\mu(ggF)$  & $1.1^{+1.3}_{-1.0}$ & 125 
 & 100\% & - & - & - & 0.85 \\
$\mu(VBF+VH)$ &  $1.6^{+0.8}_{-0.7}$ & 125 & - 
  & 59.6\% & 40.4\% & - &   \\
\hline
\multicolumn{8}{c}
{CMS (4.9${\rm fb}^{-1}$ at 7TeV + 19.7${\rm fb}^{-1}$ at 8TeV) Fig.16 of \cite{cms_tau_2014} (Jan. 2014)}\\
\hline
0 jet &  $0.34 \pm 1.09$ & 125 & 
$96.9\%$ & $1.0\%$ & $2.1\%$ & - & 0.37  \\
1 jet &  $1.07 \pm 0.46$ & 125 & 
$75.7\%$ & $14.0\%$ & $10.3\%$ & - & 0.02  \\
VBF tag &  $0.94 \pm 0.41$ & 125 & $19.6\%$ 
 & $80.4\%$ & - & - & 0.02  \\
VH tag &  $-0.33 \pm 1.02$ & 125 & - & - & 100\% &
  - & 1.70  \\
\hline
&&&&&&&subtot: 2.96 
\end{tabular}
\end{ruledtabular}
\end{table}


\begin{table}[th!]
\caption{\small \label{bestfit}
The best fitted values and the $1\sigma$ errors for the parameters in
various CP conserving fits and the corresponding chi-square per degree
of freedom and the $p$-value after the ICHEP 2014. For the SM, we
obtain $\chi^2 = 16.76$, $\chi^2/dof = 16.76/29$,  
and $p$-value = $0.966$.
}
\begin{ruledtabular}
\begin{tabular}{c|ccccc}
Cases      & {\bf CPC 1} & {\bf CPC 2} & {\bf CPC 3} & {\bf CPC 4} &          
              {\bf CPC 6} \\
\hline
            & Vary $\Delta \Gamma_{\rm tot}$ 
            & Vary $\Delta S^\gamma$, 
            & Vary $\Delta S^\gamma$, 
            & Vary $C_u^S$, $C_d^S$, 
    & Vary $C_u^S$, $C_d^S$, $C_\ell^S$, $C_v$\\
Parameters  & & $\Delta S^g$ & $\Delta S^g$, $\Delta \Gamma_{\rm tot}$
   & $C_\ell^S$, $C_v$ &  $\Delta S^\gamma$, $\Delta S^g$ \\
\hline
\multicolumn{6}{c}{After ICHEP 2014} \\
\hline
  $C_u^S$     & 1 & 1 & 1 & $0.92^{+0.15}_{-0.13}$ & $1.22^{+0.32}_{-0.38}$ \\
  $C_d^S$     & 1 & 1 & 1 & $-1.00^{+0.29}_{-0.30}$ & $-0.97^{+0.30}_{-0.34}$ \\
  $C_\ell^S$ & 1 & 1 & 1 & $0.99^{+0.17}_{-0.17}$ & $1.00^{+0.18}_{-0.17}$ \\
  $C_v$       & 1 & 1 & 1 & $0.98^{+0.10}_{-0.11}$ & $0.94^{+0.11}_{-0.12}$ \\
$\Delta S^\gamma$& 0 & $-0.72^{+0.76}_{-0.74}$ & $-0.84^{+0.80}_{-0.82}$ & 0 & $ -1.43^{+1.02}_{-0.95}$ \\
$\Delta S^g$   & 0 & $-0.009^{+0.047}_{-0.048}$ & $0.02^{+0.10}_{-0.08}$ & 0 & $-0.22^{+0.28}_{-0.24}$ \\
$\Delta \Gamma_{\rm tot}$ (MeV) & $ -0.020^{+0.45}_{-0.37}$& 0 &
$0.39^{+1.13}_{-0.76}$ & 0 &  0 \\
\hline
 $\chi^2/dof$ &$16.76/28$ & $15.81/27$ & $15.59/26$ & $16.70/25$ & $14.83/23$ \\
$p$-value & $0.953$ & $0.956$ & $0.945$ & $0.892$ & $0.901$ 
\end{tabular}
\begin{tabular}{c|cc|cccc}
Cases      & {\bf CPC N2} & {\bf CPC N3} 
           & \multicolumn{4}{c}{{\bf CPC N4}} \\
\hline
Parameters  & Vary $C_u^S$, $C_v$ 
            & Vary $C_u^S$, $C_v$, $\Delta S^\gamma$ 
            & \multicolumn{4}{c}{Vary $C_u^S$, $C_v$, $\Delta S^\gamma$, $\Delta S^g$} \\
\hline
\multicolumn{7}{c}{After ICHEP 2014} \\
\hline
  $C_u^S$     & $1.017^{+0.092}_{-0.084}$ & $1.04^{+0.10}_{-0.089}$ & $1.22^{+0.32}_{-0.37}$ & $1.22^{+0.32}_{-0.37}$ & $-1.22^{+0.37}_{-0.32}$ & $-1.22^{+0.37}_{-0.32}$ \\
  $C_d^S$     & 1 & 1 & 1 & 1 & 1 & 1  \\
  $C_\ell^S$ & 1 & 1 & 1 & 1 & 1 & 1  \\
  $C_v$       & $0.993^{+0.062}_{-0.068}$ & $0.933^{+0.078}_{-0.082}$ & $0.944^{+0.080}_{-0.084}$ & $0.944^{+0.080}_{-0.084}$ & $0.944^{+0.080}_{-0.084}$ & $0.944^{+0.080}_{-0.084}$ \\
$\Delta S^\gamma$ & 0 & $-1.10^{+0.87}_{-0.81}$ & $-1.38^{+1.02}_{-0.94}$ & $-1.38^{+1.02}_{-0.94}$ & $2.89^{+1.10}_{-1.10}$ & $2.89^{+1.11}_{-1.10}$ \\
$\Delta S^g$   & 0 & 0 & $-0.13^{+0.26}_{-0.22}$ & $-1.47^{+0.27}_{-0.24}$ & $0.20^{+0.22}_{-0.26}$ & $1.55^{+0.24}_{-0.27}$ \\
$\Delta \Gamma_{\rm tot}$ (MeV) & 0 & 0 &
0 & 0 & 0 & 0 \\
\hline
 $\chi^2/dof$ &$16.72/27$ & $15.13/26$ & $14.85/25$ & $14.85/25$ & $14.85/25$ & $14.85/25$  \\
$p$-value & $0.938$ & $0.955$ & $0.945$ & $0.945$ & $0.945$ & $0.945$ 
\end{tabular}
\end{ruledtabular}
\end{table}

\begin{table}[th!]
\caption{\small \label{cbestfit}
The best fitted values and the $1\sigma$ errors for the parameters in
the CP-violating fits and the corresponding chi-square after ICHEP 2014.
}
\begin{ruledtabular}
\begin{tabular}{c|cc|cc }
Cases & \multicolumn{2}{c|}{{\bf CPV 3}} & \multicolumn{2}{c}{{\bf CPV 4}}  \\
\hline
Parameters & \multicolumn{2}{c|}{Vary $C_u^S$, $C_u^P$, $C_v$} 
 & \multicolumn{2}{c}{Vary $\Delta S^\gamma$, $\Delta S^g$,}
        \\
        & & & \multicolumn{2}{c}{$\Delta P^\gamma$, $\Delta P^g$}  \\
\hline
\multicolumn{5}{c}{After ICHEP 2014}\\
\hline
  $C_u^S$  & $0.71^{+0.36}_{-0.46}$ & $0.71^{+0.36}_{-0.46}$ & 1 & 1  \\
  $C_d^S$   & 1 & 1 & 1 & 1\\
  $C_\ell^S$ & 1 & 1 & 1 & 1  \\
  $C_v$    & $0.946^{+0.084}_{-0.089}$  & $0.946^{+0.084}_{-0.089}$  
           & 1 & 1   \\
  $\Delta S^\gamma$ & 0 & 0 & $0.04^{+14.69}_{-1.50}$
  &  $7.90^{+6.84}_{-9.36}$   \\
  $\Delta S^g$  & 0 & 0  
  & $-0.01^{+0.05}_{-1.33}$ & $-0.38^{+0.42}_{-0.96}$  \\
$\Delta \Gamma_{\rm tot}$ (MeV) & 0 & 0 & 0 & 0  \\
\hline
$C_u^P$  &  $0.48^{+0.17}_{-1.14}$ & $-0.48^{+1.14}_{-0.17}$   & 0 & 0  \\
$\Delta P^\gamma$ & 0 & 0 
&  $-3.27^{+11.36}_{-4.83}$ & $-7.25^{+15.35}_{-0.85}$  \\
$\Delta P^g$ & 0  &0  
& $0.00^{+0.69}_{-0.69}$ & $-0.58^{+1.27}_{-0.11}$ \\
\hline
 $\chi^2/dof$ &$16.03/26$ & $16.03/26$ & $15.81/25$ & $15.81/25$  \\
$p$-value & $0.935$ & $0.935$ &  $0.920$ & $0.920$ 
\end{tabular}
\end{ruledtabular}
\end{table}

\begin{table}[th!]
\caption{\small \label{cbestfitn}
The best fitted values and the $1\sigma$ errors for the parameters in
the CP-violating fits and the corresponding chi-square after ICHEP 2014.
}
\begin{ruledtabular}
\begin{tabular}{c|cccc }
Cases & {\bf CPV N4} & {\bf CPV N5} & {\bf CPV N6} & {\bf CPV N7} \\
\hline
Parameters & Vary $C_u^S$, $C_u^P$, & Vary $C_u^S$, $C_u^P$, $C_v$, 
 & Vary $C_u^S$, $C_u^P$, $C_v$,
 & Vary $C_u^S$, $C_u^P$, $C_v$,
        \\
      & $C_v$, $\Delta S^\gamma$  & $\Delta S^\gamma$, $\Delta P^\gamma$ & $\Delta S^\gamma$, $\Delta P^\gamma$, $\Delta S^g$ & $\Delta S^\gamma$, $\Delta P^\gamma$, $\Delta S^g$, $\Delta P^g$   \\
\hline
\multicolumn{5}{c}{After ICHEP 2014}\\
\hline
  $C_u^S$ & $1.04^{+0.10}_{-0.47}$ & $1.04^{+0.10}_{-0.47}$ & $1.22^{+0.32}_{-0.67}$ & $1.20^{+0.33}_{-2.74}$ \\
  $C_d^S$ & 1 & 1 & 1 & 1 \\
  $C_\ell^S$ & 1 & 1 & 1 & 1  \\
  $C_v$   & $0.933^{+0.078}_{-0.082}$ & $0.933^{+0.078}_{-0.082}$  & $0.944^{+0.080}_{-0.084}$  
           & $0.944^{+0.080}_{-0.084}$   \\
  $\Delta S^\gamma$ & $-1.10^{+1.19}_{-0.81}$ & $-0.34^{+14.76}_{-1.57}$ & $-0.44^{+15.04}_{-1.89}$ &  $-0.38^{+18.92}_{-1.95}$ \\
  $\Delta S^g$ & 0 & 0 & $-0.13^{+0.26}_{-1.58}$  & $-0.12^{+1.91}_{-1.59}$  \\
$\Delta \Gamma_{\rm tot}$ (MeV) & 0 & 0 & 0 & 0 \\
\hline
$C_u^P$ & $0.00 \pm {0.56}$ &  $0.00 \pm {0.56}$ & $-0.07^{+0.78}_{-0.64}$ & $-0.20^{+1.74}_{-1.33}$ \\
$\Delta P^\gamma$ & 0 & $-3.19^{+11.66}_{-5.64}$ & $-3.37^{+13.03}_{-6.30}$ & $-3.06^{+14.75}_{-8.63}$  \\
$\Delta P^g$ & 0 & 0 & 0  & $0.26^{+2.03}_{-2.56}$  \\
\hline
 $\chi^2/dof$ & $15.13/25$ &$15.13/24$ & $14.85/23$ & $14.85/22$  \\
$p$-value & $0.938$ & $0.917$ & $0.900$ &  $0.869$ 
\end{tabular}
\end{ruledtabular}
\end{table}

\begin{table}[th!]
\caption{\small \label{bestfit6}
Degenerate chi-square minima of {\bf CPC 6} case.
}
\begin{ruledtabular}
\begin{tabular}{c|cccc|cccc}
Cases      & \multicolumn{8}{c}{\bf CPC 6} \\
\hline
Parameters  & \multicolumn{8}{c}{Vary $C_u^S$, $C_d^S$, $C_\ell^S$, $C_v$, $\Delta S^\gamma$, $\Delta S^g$} \\
\hline
\multicolumn{9}{c}{After ICHEP 2014} \\
\hline
  $C_u^S$     & $1.22^{+0.32}_{-0.38}$ & $1.22^{+0.32}_{-0.38}$ 
              & $1.22^{+0.32}_{-0.38}$ & $1.22^{+0.32}_{-0.38}$ 
              & $1.22^{+0.32}_{-0.38}$ & $1.22^{+0.32}_{-0.38}$ 
              & $1.22^{+0.32}_{-0.38}$ & $1.22^{+0.32}_{-0.38}$ \\
  $C_d^S$   & $0.97^{+0.34}_{-0.30}$ & $0.97^{+0.34}_{-0.30}$ 
            & $-0.97^{+0.30}_{-0.34}$ & $-0.97^{+0.30}_{-0.34}$ 
            & $0.97^{+0.34}_{-0.30}$ & $0.97^{+0.34}_{-0.30}$ 
            & $-0.97^{+0.30}_{-0.34}$ & $-0.97^{+0.30}_{-0.34}$ \\
  $C_\ell^S$ & $1.00^{+0.18}_{-0.17}$ & $-1.00^{+0.17}_{-0.18}$ 
             & $1.00^{+0.18}_{-0.17}$ & $-1.00^{+0.17}_{-0.18}$ 
             & $1.00^{+0.18}_{-0.17}$ & $-1.00^{+0.17}_{-0.18}$ 
             & $1.00^{+0.18}_{-0.17}$ & $-1.00^{+0.17}_{-0.18}$ \\
  $C_v$       & $0.94^{+0.11}_{-0.12}$ & $0.94^{+0.11}_{-0.12}$ 
              & $0.94^{+0.11}_{-0.12}$ & $0.94^{+0.11}_{-0.12}$ 
              & $0.94^{+0.11}_{-0.12}$ & $0.94^{+0.11}_{-0.12}$ 
              & $0.94^{+0.11}_{-0.12}$ & $0.94^{+0.11}_{-0.12}$ \\
$\Delta S^\gamma$ & $-1.39^{+1.02}_{-0.95}$ & $-1.44^{+1.02}_{-0.95}$ 
                  & $-1.43^{+1.02}_{-0.95}$ & $-1.47^{+1.02}_{-0.95}$ 
                  & $-1.40^{+1.02}_{-0.95}$ & $-1.44^{+1.02}_{-0.95}$ 
                  & $-1.43^{+1.02}_{-0.95}$ & $-1.47^{+1.02}_{-0.95}$\\
$\Delta S^g$   & $-0.14^{+0.29}_{-0.25}$ & $-0.14^{+0.29}_{-0.25}$ 
               & $-0.22^{+0.28}_{-0.24}$ & $-0.22^{+0.28}_{-0.24}$ 
               & $-1.47^{+0.28}_{-0.24}$ & $-1.47^{+0.28}_{-0.24}$ 
               & $-1.54^{+0.28}_{-0.25}$ & $-1.54^{+0.28}_{-0.25}$ \\
$\Delta \Gamma_{\rm tot}$ (MeV) & 0 & 0 &
0 &0&  0 &
0 &0&  0\\
\hline
 $\chi^2/dof$ &$14.83/23$ & $14.83/23$ & $14.83/23$ & $14.83/23$ & $14.83/23$ & $14.83/23$ & $14.83/23$ & $14.83/23$ \\
$p$-value & $0.901$ & $0.901$ & $0.901$ & $0.901$ & $0.901$ & $0.901$ & $0.901$ & $0.901$ \\
\hline
\hline 
Cases      & \multicolumn{8}{c}{\bf CPC 6} \\
\hline
Parameters  & \multicolumn{8}{c}{Vary $C_u^S$, $C_d^S$, $C_\ell^S$, $C_v$, $\Delta S^\gamma$, $\Delta S^g$} \\
\hline
\multicolumn{9}{c}{After ICHEP 2014} \\
\hline
  $C_u^S$     & $-1.23^{+0.38}_{-0.32}$ & $-1.23^{+0.38}_{-0.32}$ 
              & $-1.23^{+0.38}_{-0.32}$ & $-1.23^{+0.38}_{-0.32}$ 
              & $-1.23^{+0.38}_{-0.32}$ & $-1.23^{+0.38}_{-0.32}$ 
              & $-1.23^{+0.38}_{-0.32}$ & $-1.23^{+0.38}_{-0.32}$ \\
  $C_d^S$   & $0.97^{+0.34}_{-0.30}$ & $0.97^{+0.34}_{-0.30}$ 
            & $-0.97^{+0.30}_{-0.34}$ & $-0.97^{+0.30}_{-0.34}$ 
            & $0.97^{+0.34}_{-0.30}$ & $0.97^{+0.34}_{-0.30}$ 
            & $-0.97^{+0.30}_{-0.34}$ & $-0.97^{+0.30}_{-0.34}$ \\
  $C_\ell^S$ & $1.00^{+0.18}_{-0.17}$ & $-1.00^{+0.17}_{-0.18}$ 
             & $1.00^{+0.18}_{-0.17}$ & $-1.00^{+0.17}_{-0.18}$ 
             & $1.00^{+0.18}_{-0.17}$ & $-1.00^{+0.17}_{-0.18}$ 
             & $1.00^{+0.18}_{-0.17}$ & $-1.00^{+0.17}_{-0.18}$ \\
  $C_v$       & $0.94^{+0.11}_{-0.12}$ & $0.94^{+0.11}_{-0.12}$ 
              & $0.94^{+0.11}_{-0.12}$ & $0.94^{+0.11}_{-0.12}$ 
              & $0.94^{+0.11}_{-0.12}$ & $0.94^{+0.11}_{-0.12}$ 
              & $0.94^{+0.11}_{-0.12}$ & $0.94^{+0.11}_{-0.12}$ \\
$\Delta S^\gamma$ & $2.90^{+1.11}_{-1.11}$ & $2.85^{+1.11}_{-1.11}$ 
                  & $2.87^{+1.12}_{-1.11}$ & $2.82^{+1.12}_{-1.11}$ 
                  & $2.90^{+1.11}_{-1.11}$ & $2.85^{+1.11}_{-1.11}$ 
                  & $2.87^{+1.12}_{-1.11}$ & $2.82^{+1.11}_{-1.11}$\\
$\Delta S^g$   & $0.22^{+0.24}_{-0.28}$ & $0.22^{+0.24}_{-0.28}$ 
               & $0.14^{+0.25}_{-0.29}$ & $0.14^{+0.25}_{-0.29}$ 
               & $1.54^{+0.25}_{-0.28}$ & $1.54^{+0.25}_{-0.28}$ 
               & $1.47^{+0.24}_{-0.28}$ & $1.47^{+0.24}_{-0.28}$ \\
$\Delta \Gamma_{\rm tot}$ (MeV) & 0 & 0 &
0 &0&  0 &
0 &0&  0\\
\hline
 $\chi^2/dof$ &$14.83/23$ & $14.83/23$ & $14.83/23$ & $14.83/23$ & $14.83/23$ & $14.83/23$ & $14.83/23$ & $14.83/23$ \\
$p$-value & $0.901$ & $0.901$ & $0.901$ & $0.901$ & $0.901$ & $0.901$ & $0.901$ & $0.901$
\end{tabular}
\end{ruledtabular}
\end{table}

\end{document}